\documentclass[linenumbers]{aastex631}

\shorttitle{Spectroscopic Confirmation of Two Galaxy Clusters}
\shortauthors{Chen et al.}
\graphicspath{{./}{figures/}}
\usepackage{array}
\usepackage{soul}
\usepackage{tablefootnote}
\usepackage{colortbl}  %彩色表格需要加载的宏包
\usepackage{xcolor}

\begin{document}
\nolinenumbers
\title{Spectroscopic Confirmation of Two X-ray Diffuse and Massive Galaxy Clusters at Low Redshift}

\author[0000-0003-3536-5504]{Kaiyuan Chen}
\affiliation{Department of Astronomy, School of Physics, Peking University, Beijing 100871, China}

\author[0000-0002-9587-6683]{Weiwei Xu}
\affiliation{Kavli Institute for Astronomy and Astrophysics, Peking University, Beijing 100871, China}

\author[0000-0003-4176-6486]{Linhua Jiang}
\affiliation{Department of Astronomy, School of Physics, Peking University, Beijing 100871, China}
\affiliation{Kavli Institute for Astronomy and Astrophysics, Peking University, Beijing 100871, China}

\begin{abstract}
\nolinenumbers

We present MMT spectroscopic observations of two massive galaxy cluster candidates at redshift $z\sim0.07$ that show extended and diffuse X-ray emission in the ROSAT All Sky Survey (RASS) images. The targets were selected from a previous catalog of 303 newly-identified cluster candidates with the similar properties using the intra-cluster medium emission. Using the new MMT Hectospec data and SDSS archival spectra, we identify a number of member galaxies for the two targets and confirm that they are galaxy clusters at $z=0.079$ and 0.067, respectively. The size of the two clusters, calculated from the distribution of the member galaxies, is roughly 2 Mpc in radius. 
{We estimate cluster masses using three methods based on their galaxy number overdensities, galaxy velocity dispersions, and X-ray emission. The overdensity-based masses are $(6\sim8) \rm \times10^{14}\ M_\odot$, comparable to the masses of large clusters at low redshift. The masses derived from velocity dispersions are significantly lower, likely due to their diffuse and low concentration features.}
Our result suggests the existence of a population of large clusters with very diffuse X-ray emission that have been missed by most previous searches using the RASS images. If most of the 303 candidates in the previous catalog are confirmed to be real clusters, this {may help to reduce the discrepancy of cosmological results} between the CMB and galaxy cluster measurements.

\end{abstract}


\section{Introduction}        
\label{sect:intro}

Galaxy clusters are the largest gravitationally-bounded systems in the universe.
They are astrophysical laboratories for studying gravitational lensing, dark matter, galaxy properties and evolution in dense environments \citep{d_1980,bo_1984,Tyson1990,m_1996,Spergel2000,Seljak2000,lm_2004,sand_2004,kn_2011,kab_2012}. They are also excellent tracers of the cosmic large-scale structure, and provide an independent and powerful tool  to test and constrain cosmological models. This tool is complementary to other methods \citep[e.g.,][]{a_2011}, such as cosmic microwave background (CMB) \citep[e.g.,][]{d_2009}. However, there is a tension between the cosmological constraints obtained from CMB and from galaxy clusters. Based on the cosmological parameters derived from CMB, more galaxy clusters are expected than those detected in the current data \citep{lambda,chandra_cosmo}. This tension has been reinforced recently by the Planck results \citep{planck}. 

Different methods have been used to search for galaxy clusters \citep{Cavaliere1976,Lopes2004}. One major method is to identify member galaxies in the optical and/or near-IR band \citep{dress,n2021}. In this approach, one first identifies galaxy overdense regions using spectroscopic or photometric redshifts of galaxies, and then choose the regions that satisfy the density and size criteria of galaxy clusters. A large number of galaxy clusters have been discovered using this method \citep{abell89,Wen2012,Wen2018,Zou2021}. Another major method is based on X-ray images \citep{chandra}. The X-ray radiation is produced by hot intra-cluster medium (ICM) in virialized clusters. It is an efficient way to find galaxy clusters when it is combined with optical/near-IR data. A number of catalogs of clusters have been constructed using X-ray data \citep{gio,Bohringer2000,Bohringer2004,Bohringer2014,Bohringer2017,Piffaretti2011,kirk}. Many of them were made with the ROSAT data, including ROSAT-ESO Flux Limited X-ray Galaxy Cluster Survey \citep{Bohringer2004,Bohringer2014}, Northern ROSAT All-Sky Galaxy Cluster Survey \citep{Bohringer2000,Bohringer2017}, ROSAT Brightest Cluster Sample \citep{Ebeling2000}, a Catalog of Clusters of Galaxies in a Region of 1 steradian around the South Galactic Pole \citep{Cruddace2002} and the ROSAT North Ecliptic Pole survey \citep{Henry2006}). {Most of} these catalogs have been compiled in the Meta-Catalog of X-ray detected Clusters \citep{Piffaretti2011}. 

Recently, \cite{xu_2018,xu_2021} re-analyzed the ROSAT All Sky Survey (RASS) images using a combination of wavelet filtering, source extraction, and maximum likelihood fitting. They identified 944 very extended galaxy clusters, and constructed an X-ray selected catalog of extended clusters from the ROSAT All-Sky Survey (referred to as the RXGCC catalog). A number of clusters in this catalog are newly identified with ICM emission, but their nature remains unclear. {They exhibit flatter surface-brightness distributions and weaker X-ray emission than regular clusters.}
If these candidates are real clusters, they {may help to reduce} the tension between the CMB and galaxy cluster results mentioned above. In this work, we present our spectroscopic observations of two extended cluster candidates selected from the RXGCC catalog using the MMT Hectospec spectrograph. Our purpose was to spectroscopically confirm the two clusters and measure their basic properties. We obtained more than 300 galaxy spectra from the MMT. The combination of the MMT spectra and archival Sloan Digital Sky Survey (SDSS) spectra allowed us to confirm that both candidates are galaxy clusters.

The structure of the paper is as follows. In Section 2, we briefly introduce our target selection, the MMT observations, and data reduction. In Section 3, we show our observational results. We discuss and summarize our results in Section 4. Throughout the paper, we use a $\Lambda$CDM cosmology with $\rm \Omega_m =  0.3$, $\Omega_\Lambda = 0.7$, and $\rm H_0 = 70\ km\ s^{-1}\ Mpc^{-1}$, and assume the average density of the universe $\rm \bar\rho=9.9\times 10^{-30}\ g\ cm^{-3}$.

\section{Target Selection and Spectroscopic Observations}
\label{sect:Obs}

\subsection{Target Selection}

We selected our targets from the RXGCC catalog by \cite{xu_2021}. 
The basic procedure of detecting extended clusters using X-ray images in \cite{xu_2021} is summarized below. The algorithm includes a multi-resolution wavelet filtering of the RASS image at [0.5-2.0] keV, a source extraction, and a maximum likelihood fitting to characterize detections. They also cross-matched detections with previously identified galaxy clusters, combined the spatial and redshift distributions of nearby galaxies, and removed false detections by visual inspection of multiple wavelength images. The redshift and its uncertainty of each cluster were estimated by the redshift distribution of galaxies within $15\arcmin$ from the cluster center. Both spectroscopic and photometric redshifts of galaxies were considered, using data from the SDSS DR16, Galaxy And Mass Assembly (GAMA) DR1, the Two Micron All Sky Survey Photometric Redshift Catalog (2MPZ catalog, \citealt{Bilicki2014}), and the NASA Extragalactic Database (NED). As a result, 944 cluster candidates were detected, including 303 newly-identified clusters with ICM emission. Compared to galaxy clusters found with ICM emission in other studies, these 303 clusters are statistically less massive with much flatter surface brightness profiles.

\begin{figure}[]
\centering  %图片全局居中
\includegraphics[width=0.8\textwidth]{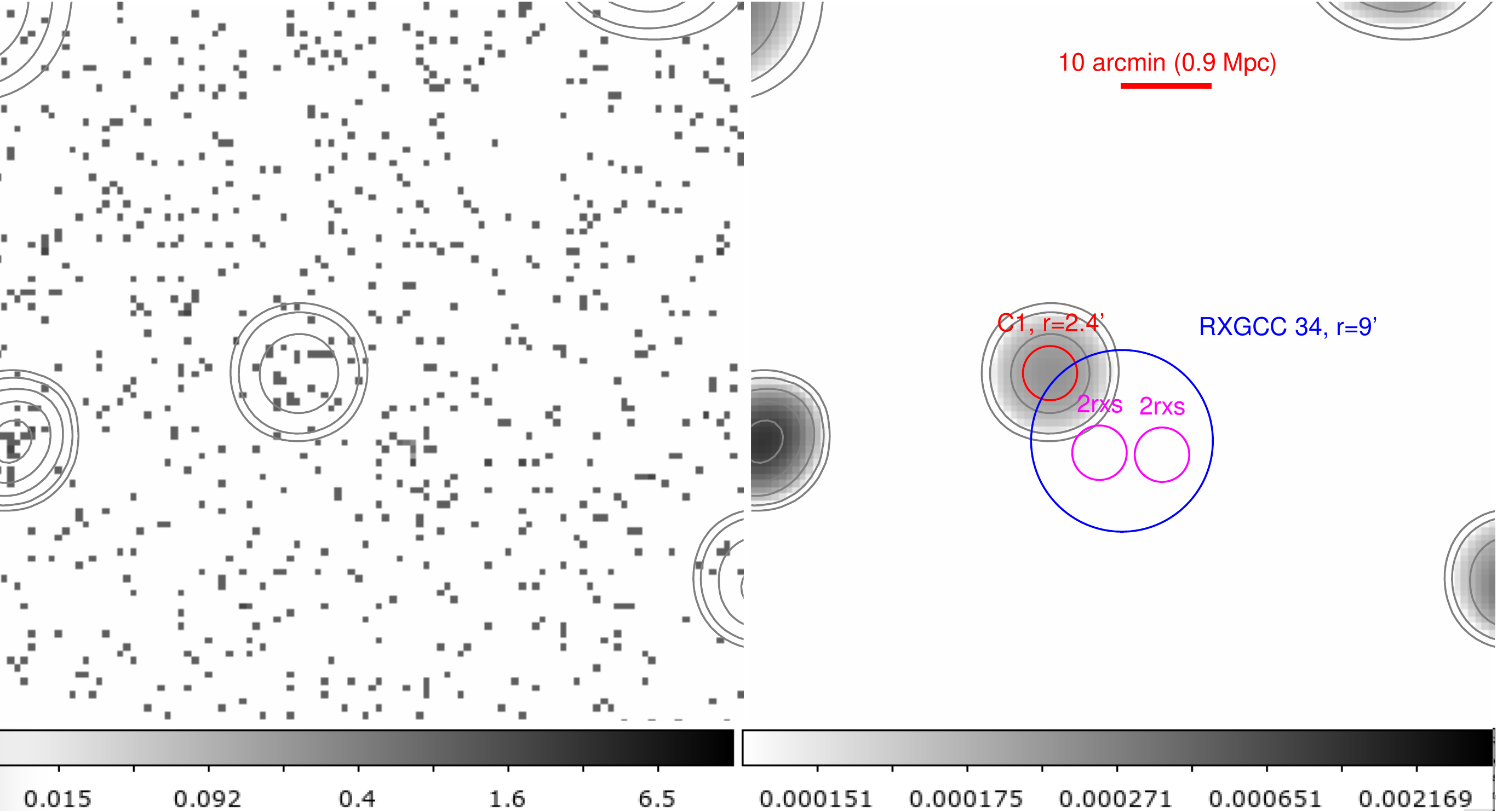}
\includegraphics[width=0.8\textwidth]{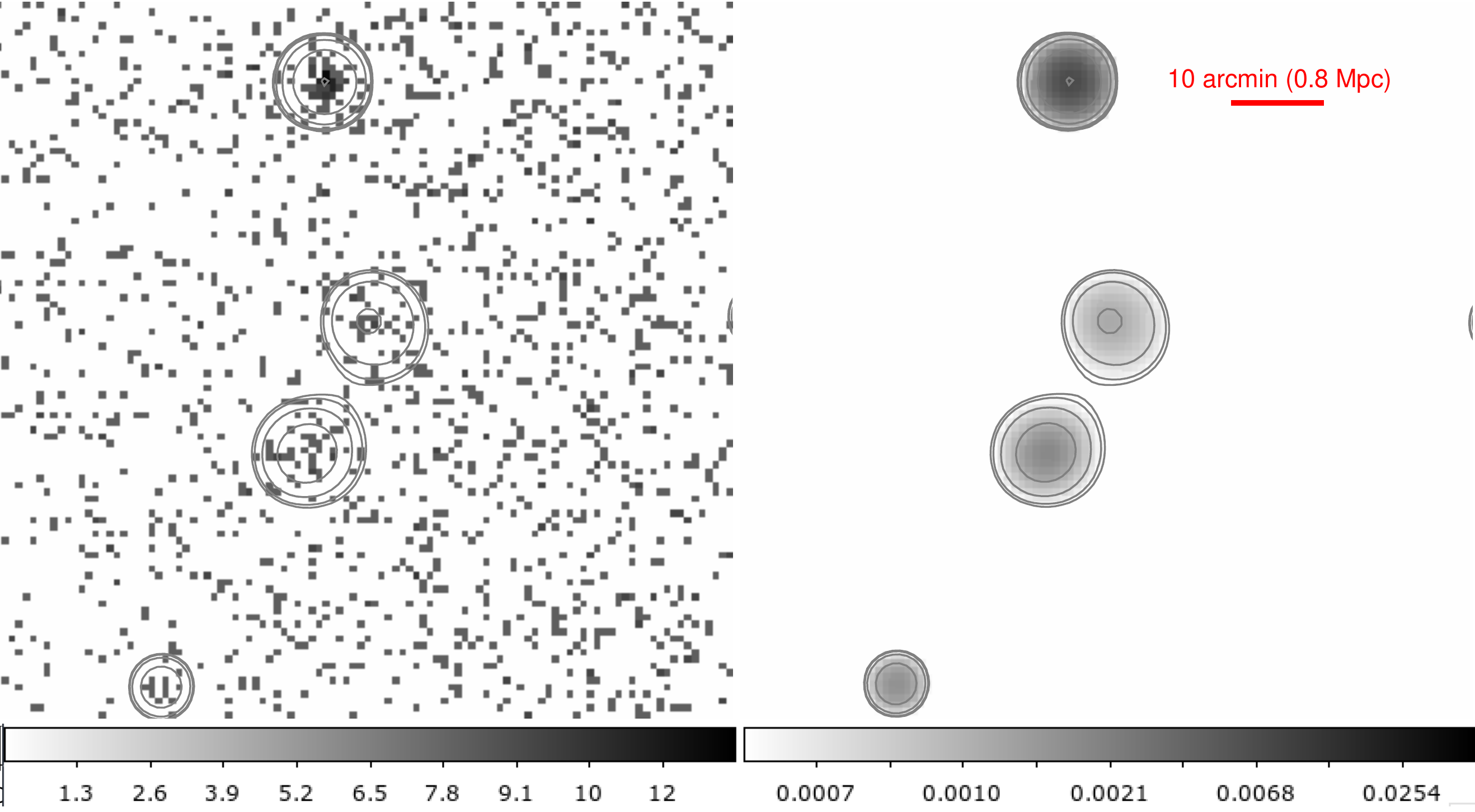}
\caption{
X-ray images of the two galaxy clusters C1 (upper panels) and C2 (lower panels). For each cluster, the left panel is the ROSAT X-ray photon image in the energy of [0.5-2.0] keV, in the logarithm scale, and the right panel is the reconstructed image after the wavelet filtering shown in logarithm scale. The grey contours have 6 logarithm levels with the lower and upper limits of 0.00014 and 0.0025 for C1, and 0.00061 and 0.0069 for C2. 
In panel for C1, we have subtracted X-ray pollution from two nearby quasars. Two magenta circles label out the quasar-contaminated regions. The area marked with orange circles is the initial detection of C1 in \cite{xu_2021}. After subtraction, the corrected size and position is shown with the red circle.
\label{Fig_X0}
}
\end{figure}

\begin{figure}[]
\centering  %图片全局居中
%\subfigure[Cluster\_1]{
\includegraphics[width=0.45\textwidth]{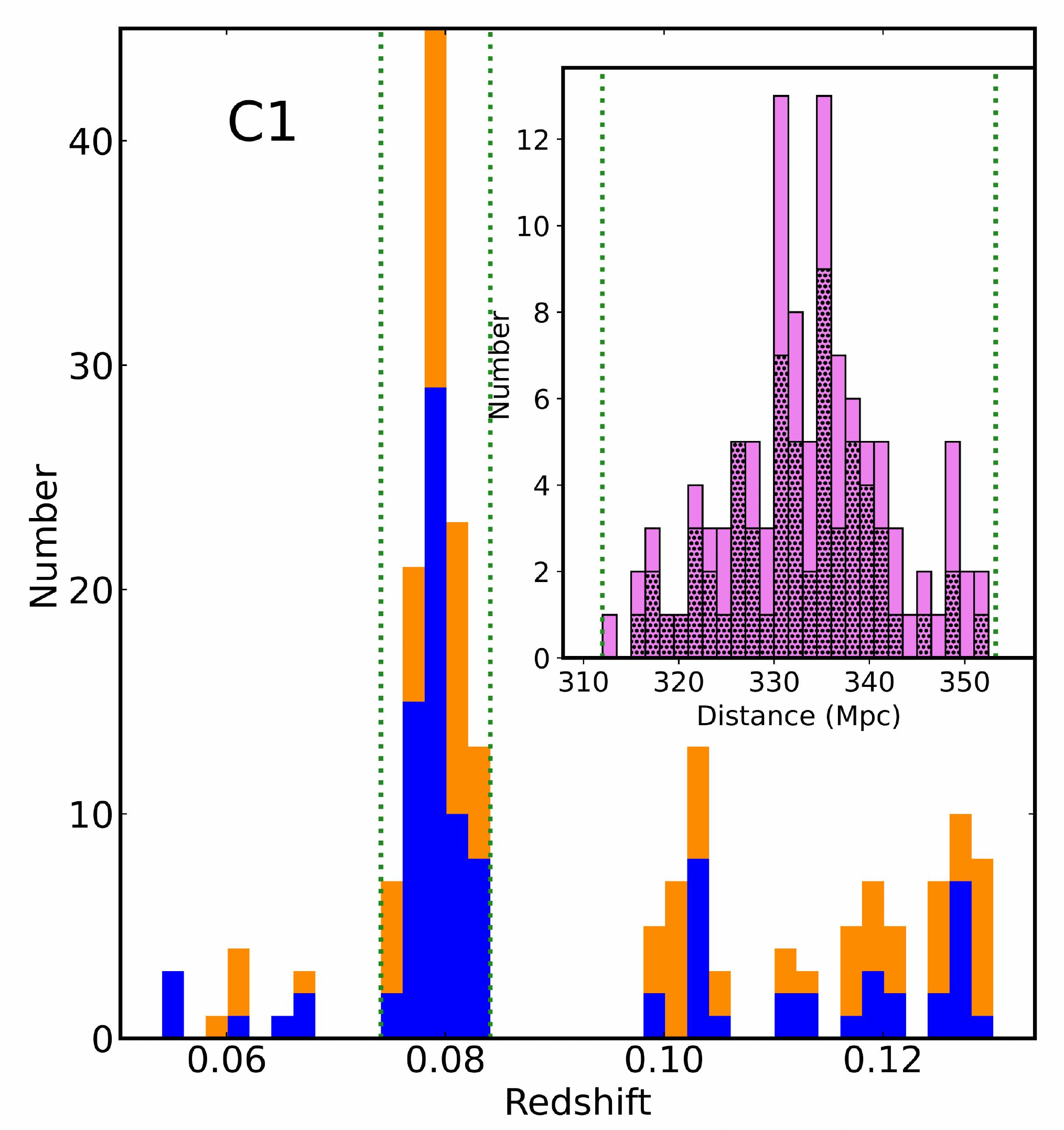}%}
%\subfigure[Cluster\_2]{
\includegraphics[width=0.45\textwidth]{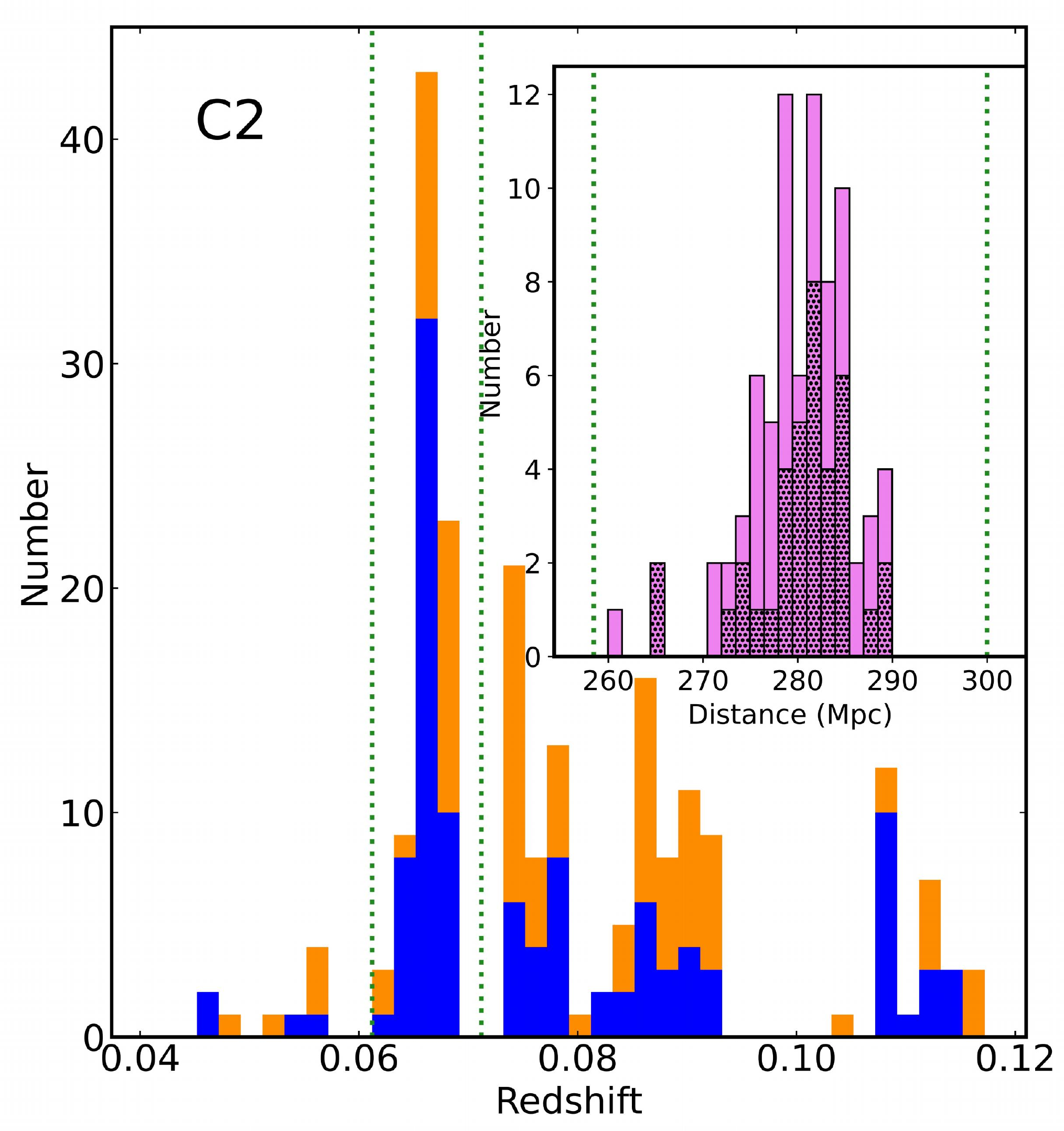}%}
\caption{Distributions of spectroscopic redshifts for galaxies within $35\arcmin$ from the cluster centers. The blue histograms represent galaxies from SDSS and the orange histograms represent galaxies from MMT. The bin size is 0.002. In each panel, the green dotted lines indicate the cluster redshift range $z_c \pm  \sigma_{zc}$ from the RXGCC catalog. The inset shows the zoomed-in distribution {of both SDSS and MMT galaxies} around the cluster redshift. The dotted area labels out the member galaxies.}
%cmt{The colors in inset should be the same with the panels.}
\label{hist}
\end{figure}

We selected two newly ICM-identified, very extended cluster candidates from the RXGCC catalog of \cite{xu_2021}. The brief selection procedure is as follows. For each of the 303 clusters in the catalog, we first identified its possible member galaxies with $r$-band magnitudes between 15.5 and 21.3 mag within $35\arcmin$ from the cluster center. We used spectroscopic and photometric redshifts available from SDSS DR16, NED, GAMA DR3, and 2MPZ. Member galaxies were required to have spectroscopic redshifts within $z_{c} \pm \sigma_{zc}$ (criterion 1) or photometric redshifts within $z_{c} \pm 2\times\sigma_{zc}$ (criterion 2), where $z_{c}$ and $\sigma_{zc}$ are the redshift of the cluster and its uncertainty. The numbers of galaxies that satisfy criteria 1 and 2 are expressed as $n_1$ and $n_2$, respectively. We further required $n_1\ge30$ and $n_2\ge 180$. From the result, we selected two clusters (denoted as C1 and C2) for our follow-up MMT observations. Table \ref{t1} lists the basic information and properties of the two clusters. Figure \ref{Fig_X0} shows the X-ray images of the two clusters. Figure \ref{hist} shows the spectroscopic redshift distributions of the galaxies around the two clusters. The blue histograms in the figure represent galaxies  from the SDSS data. These galaxies are typically bright. From the figure, C1 has a strong peak at $z\sim0.079$ in its redshift distribution, and C2 has a clear peak at $z\sim0.067$. 

The MMT Hectospec observations were used to observe photometrically selected galaxies that did not have previous spectroscopic observations. These galaxies are relatively fainter. 

\begin{deluxetable*}{c|ccc}
\tablecaption{\label{t1}Information of the two galaxy clusters}
\setlength{\tabcolsep}{15pt}
\tablewidth{0.9 \textwidth}
\tablehead{
\colhead{Method} &\colhead{Parameters} & \colhead{C1 } &  \colhead{C2 } 
}
\startdata
& {RXGCC ID.} & {34} & {302} \\
&R.A.& {00 43 06}& 08 50 10 \\
&Decl.  & {15 17 56} & 32 50 35\\
&Redshift & 0.079$\pm$0.005 & 0.066$\pm$0.005\\
X-ray &{Detection radius}&{2.4$\arcmin$} & {13.6$\arcmin$}\\
(RXGCC)&{R$_{500}$\tablenotemark{a}} & {4$\arcmin$, 0.53 Mpc} & {10$\arcmin$, 1.09 Mpc}\\
&{Hardness ratio\tablenotemark{b}} & {0.95} & {0.92}\\
&{X-ray flux ($\rm 10^{-12} erg\cdot s^{-1}\cdot cm^{-2}$)} &{0.39} &{21.04} \\
&{X-ray luminosity ($\rm 10^{44} erg\cdot s^{-1}$)} &{0.06} &{2.23}\\
&{$\rm M_X$ ($\rm 10^{14}M_\odot$)\tablenotemark{c}} & {0.46} & {3.88}\\
\hline
&R.A. & 00 42 43 & 08 50 24 \\
&Decl. & 15 13 31 & 32 50 44\\
Optical&Redshift & 0.079 & 0.067 \\
(SDSS and MMT)&{Number of }member galaxies & {87} & {52}\\
&$\rm R_{200}$ / $\rm R_{500}$ (Mpc) & {1.9 / 1.3} & {1.7 / 1.0}\\
&{M$_\delta$\tablenotemark{d} ($\rm 10^{14}M_\odot$)}& {7.9 / 5.7} & {5.7 / 2.9} \\
&{M$_\sigma$\tablenotemark{e} ($\rm 10^{14}M_\odot$)}& {2.1} & {0.7} \\
\enddata

\tablecomments{{
The information in the upper part of the table is from \cite{xu_2018,xu_2021}, whereas the lower part is from this work. Some parameters for C1 are slightly different from those in \cite{xu_2021} (see Section \ref{discuss}).}}

\tablenotetext{a}{{Radius within which the mean density is 500 times the average density at the same redshift.} }
\tablenotetext{b}{{Ratio between photon numbers detected in [0.5-2.0 keV] and [0.1-0.4 keV] within the radius of $\rm R_{500}$.}} %ww
\tablenotetext{c}{{Mass within $\rm R_{500}$ derived from the X-ray data.}} 
\tablenotetext{d}{{$\rm M_{200}$ and $\rm M_{500}$ derived from the overdensities of the member galaxies.}}
\tablenotetext{e}{{$\rm M_{200}$ estimated from the velocity dispersions of the member galaxies.}}
\end{deluxetable*}

\subsection{Observations and Data Reduction}

Spectroscopic observations of the two clusters were made with MMT Hectospec \citep{Fabricant2005} in September and October, 2020. Hectospec is a multi-fiber optical spectrograph. It has 300 fibers over a field of view of $\sim 1^\circ$ in diameter. We chose to use a 270 gpm grating, which provides a wavelength coverage of $3700\sim9200$ \AA\ with a resolving power of $\sim 1000$. This spectral resolution is sufficient to identify emission lines in our spectra.

\begin{figure}[]
\centering  %图片全局居中
\includegraphics[width=0.5\textwidth]{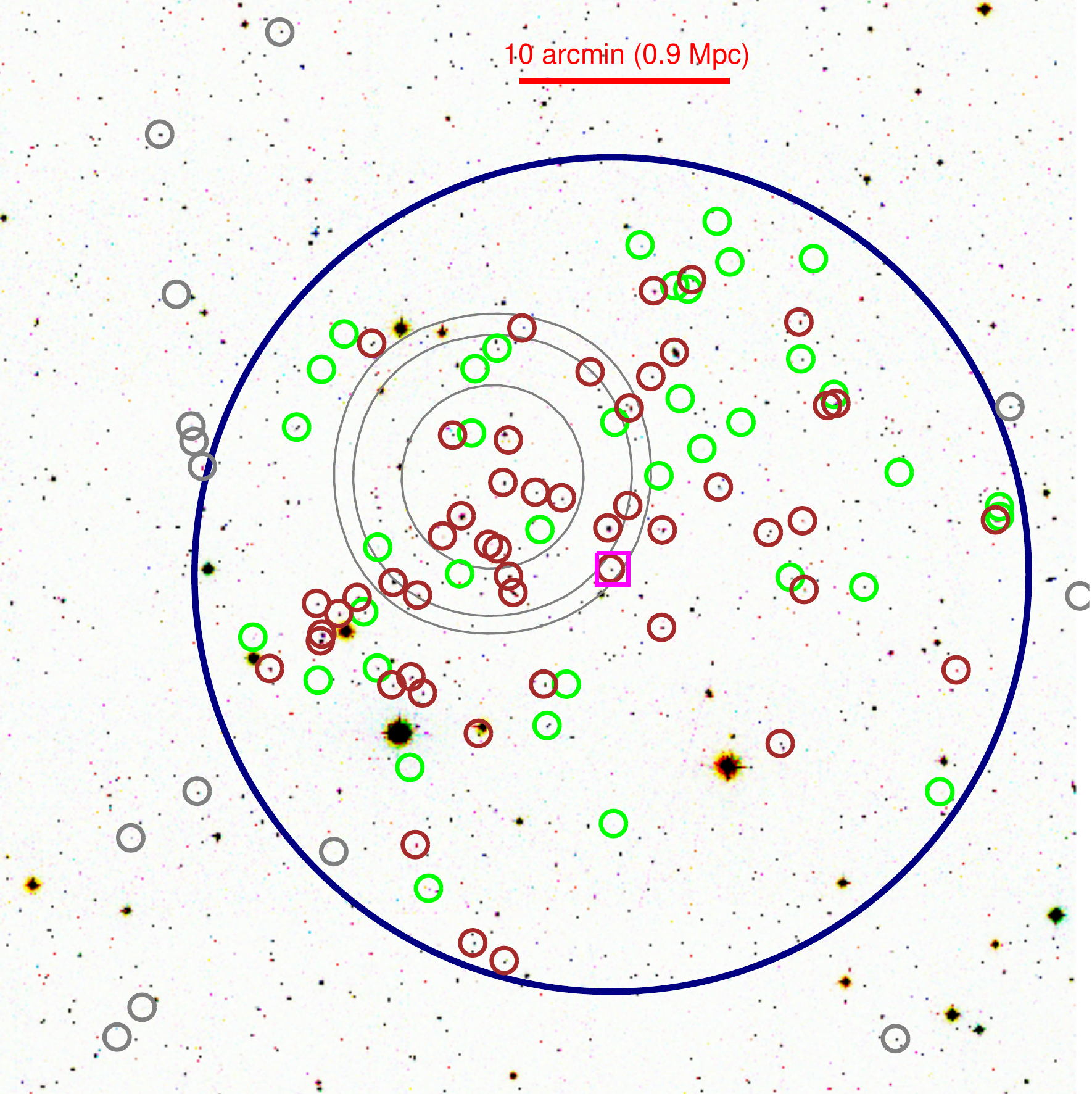}%}
\includegraphics[width=0.5\textwidth]{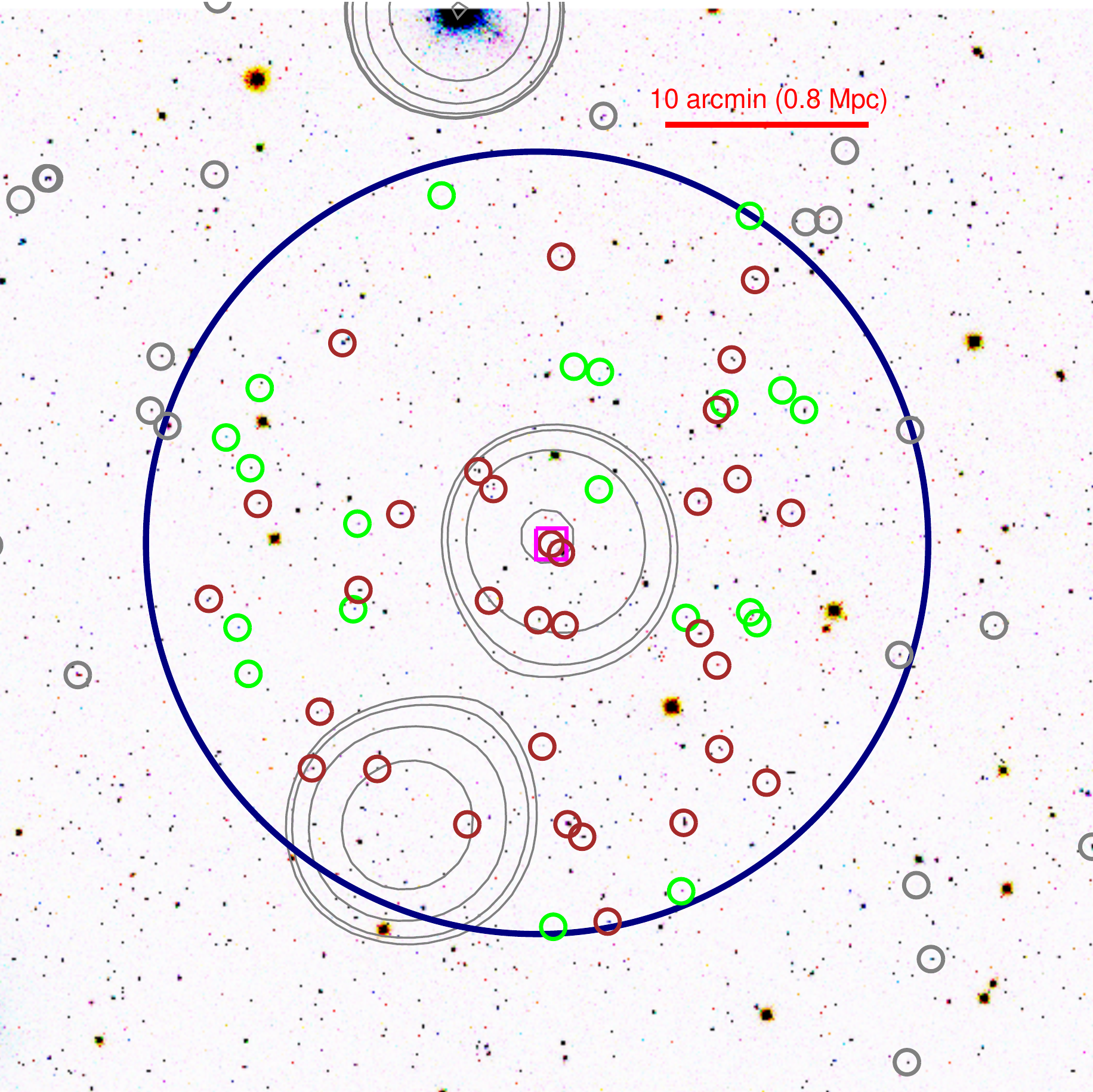}%}
\caption{\label{Fig_X}
Galaxy distributions for C1 (left panel) and C2 (right panel) {in the RGB image (combination of the $i$, $r$ and $g$ bands) from the SDSS DR12.}  {We} show the contours of the X-ray images from Figure \ref{Fig_X0}. The {grey} and color-coded circles represent galaxies between green dashed lines in Figure \ref{hist}. The brown and green circles label out the cluster member galaxies within {R$_{200}$}, and they are from the SDSS and MMT, respectively. The magenta boxes outline the centers of two clusters. {The large dark blue circles indicate the cluster regions.}}
\end{figure}

For cluster C1, we observed 168 member galaxy candidates. The total on-source integration time was 80 minutes, including 4 exposures with 20 minutes per exposure. The weather condition was good. For cluster C2, we also observed 168 member galaxy candidates. We obtained a total of 6 exposures with 20 minutes per exposure. The weather condition was moderate to poor. We discarded the worst exposure. The central regions of the two fields are shown in Figure \ref{Fig_X}.

The MMT data were reduced by the HSRED pipeline, which is an IDL package developed for the reduction of the Hectospec data. 
We first used the pipeline to de-bias and flat-field the raw images, and remove cosmic rays. We then used dome flats to remove CCD fringing and high-frequency flat-fielding variations. Sky spectra were used to construct sky emission models. These model spectra were scaled and subtracted from galaxy spectra. Finally, each galaxy spectrum was extracted. The pipeline also made the wavelength calibration by cross-correlating observed spectra against the calibration arc spectra. The data products include one dimensional (1D) variance-weighted science spectra and error spectra. We briefly evaluated the error spectra using the science spectra. We calculated the standard deviations of the continuum spectra (after removing line emission) in the wavelength range between 6563$\rm\AA$ and 8200$\rm\AA$, and compared them with the error spectra. They are generally consistent within $\pm10$\%.

\section{Results} 

\subsection{Observational Results} \label{obr}

\begin{figure}[]
\centering 
\includegraphics[width=0.8\textwidth]{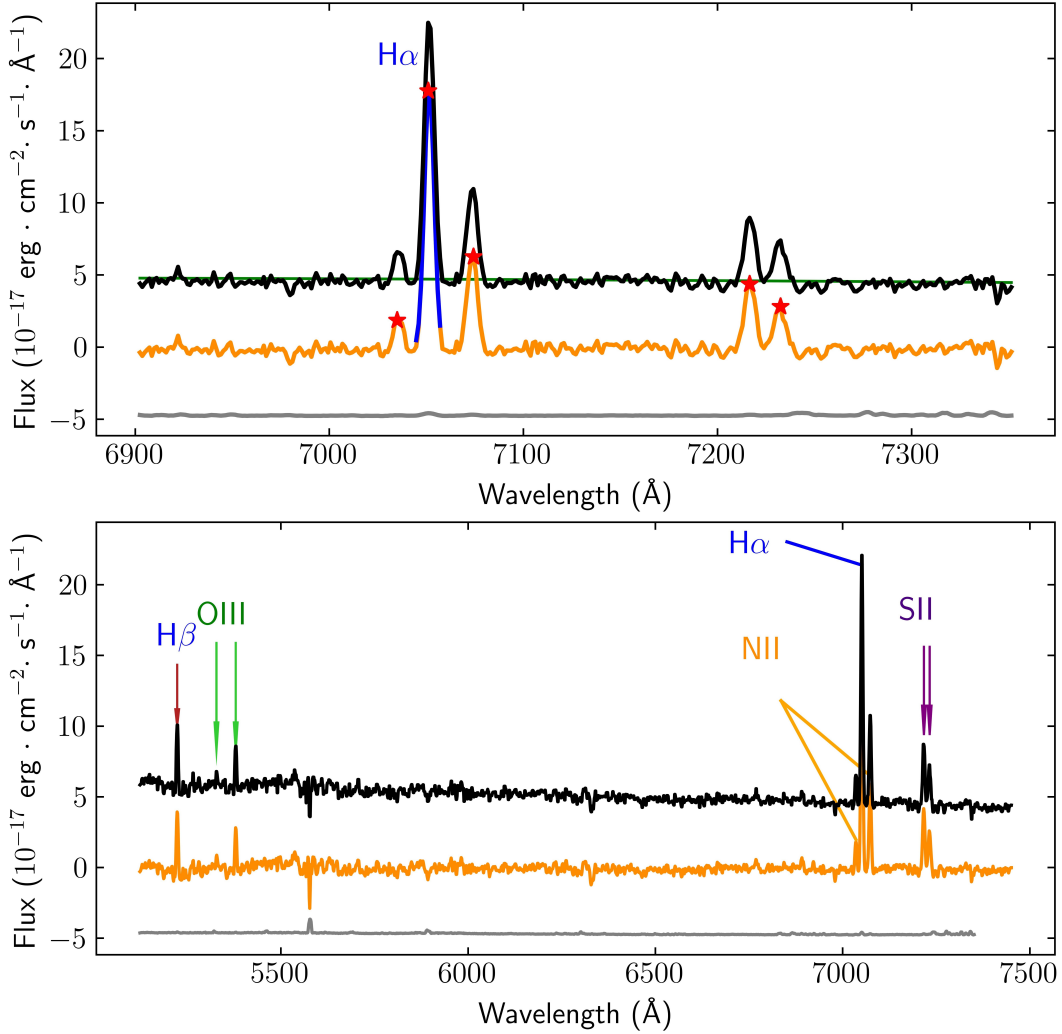}
\caption{
MMT Hectospec spectrum of one galaxy. The upper panel shows the wavelength range around the H$\alpha$ emission line and the lower panel shows a wider wavelength range. The science spectrum is in black and the continuum-subtracted spectrum is in orange. The error spectrum is in grey and has been vertically shifted by --5 for clarity. The red stars
{in the upper panel}
indicate the peaks of the identified emission lines. 
\label{Fig:spec}
}
\end{figure}

From the Hectospec observations, we obtained a total of 336 object spectra, 168 for either galaxy cluster. We first identify galaxies and measure their redshifts. The average signal-to-noise ratio (S/N) of the continuum at $r\sim21$ mag is below 2 per pixel. Most galaxies fainter than $r=21$ mag do not have sufficient S/N for us to measure reliable redshifts, if they do not have emission lines. Therefore, we use the H$\alpha$ emission line to identify galaxies (i.e., H$\alpha$ emitters) and calculate redshifts (Figure \ref{Fig:spec}). 

For each Hectospec spectrum, we first subtract its continuum. We use a cubic polynomial to fit the spectrum between 6563 $\rm \AA$  and 8200 $\rm \AA$. This wavelength range corresponds to a redshift range of 0--0.25 for the H$\alpha$ line detection. After the subtraction of the continuum, we search for the strongest emission line (presumably H$\alpha$) in this wavelength range. The line must also satisfy two criteria: 1) at least 4 contiguous pixels with S/N $>1.5$; 2) the two highest pixels with S/N $>5$. In order to measure the line flux, we use a Gaussian profile to fit the 13 nearest pixels around the line center and obtain its full width at half maximum (FWHM). The line flux and its error are calculated by summing the pixels in a range of $1.5\times$ FWHM. The redshift is calculated from the best Gaussian fitting result. Finally, we obtain a total of 271 H$\alpha$ emitters, including 127 galaxies for C1 and 144 galaxies for C2.

The orange histograms in Figure \ref{hist} show the redshift distributions of the newly identified galaxies  by the MMT observations. In either cluster, there is a prominent redshift distribution peak that is coincident with that from the SDSS spectra. Our new data, together with the SDSS data, confirm that C1 and C2 are two galaxy clusters.

\subsection{Cluster Properties}\label{ic}

We measure the properties of the galaxy clusters using the SDSS and MMT data in this section. We first estimate their physical centers and sizes. For either cluster, we first obtain its preliminary redshift coverage using the redshift distribution of the galaxies shown in Figure \ref{hist}. The inset of Figure \ref{hist} displays more detailed distribution of galaxies at its corresponding redshift range. We then locate its preliminary physical center as follows. We start with the SDSS and MMT galaxies in the same redshift range in one deg$^2$ around the X-ray center. For each of these galaxies, we count the number of its nearby galaxies within a radius of 2 Mpc, the typical size of galaxy clusters. We take the galaxy with the largest number of nearby galaxies as the preliminary center of this cluster. 

\begin{figure}[t]
\centering  %图片全局居中
%\subfigure[Cluster\_1]{
\includegraphics[width=0.8\textwidth]{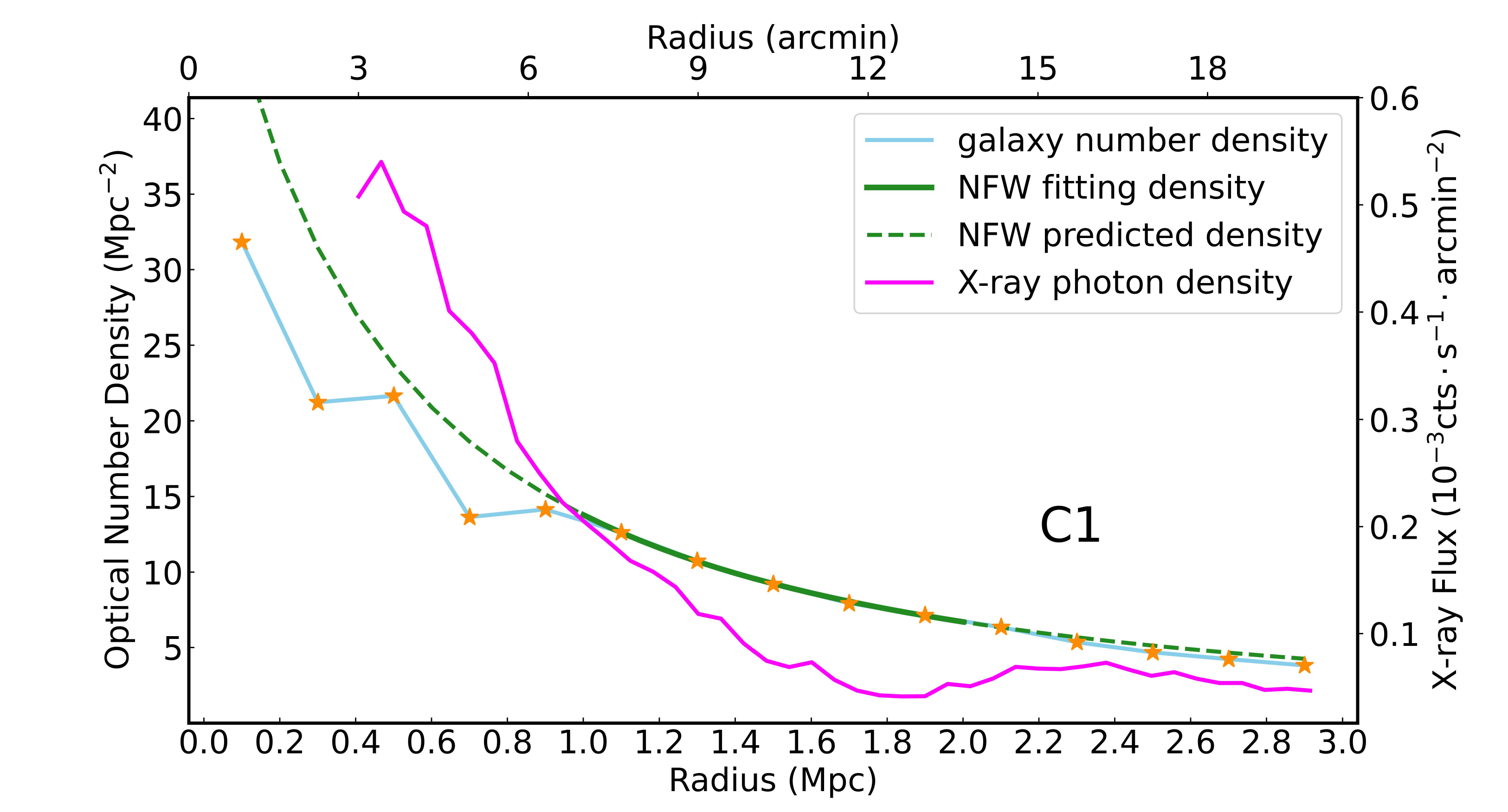}%}

%\subfigure[Cluster\_2]{
\includegraphics[width=0.8\textwidth]{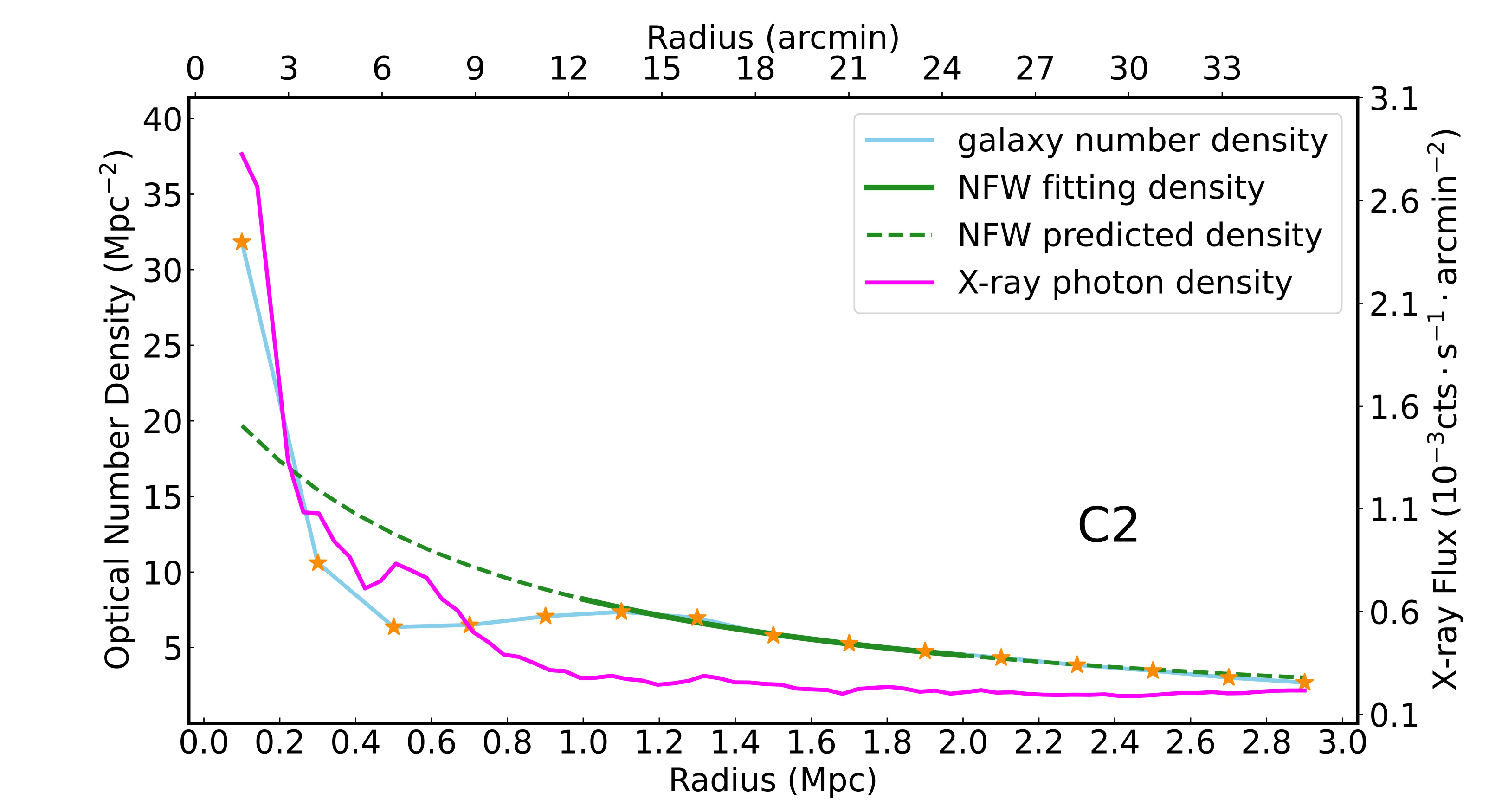}%}
\caption{
Galaxy number densities (shown as stars) as a function of radius. 
{The magenta curve shows the X-ray photon number profile (units is labelled on the right). The green solid line is the best fit of the NFW model profile using the data points between 1 and 2 Mpc. The green dashed line is prediction from the best fit.}
}
\label{fig:growth}
\end{figure}

We then determine a preliminary size of the cluster using the galaxies in the above redshift range around the above center.
{We draw circles around the center and estimate the overdensity of the galaxies within these circles. We take $\rm R_{200}$ as the radius of the cluster. The galaxy bias is also considered \citep{Crocce2016}. After obtaining the preliminary size, we repeat this procedure to refine the center and size.} The final centers of the two clusters are illustrated by the magenta boxes in Figure \ref{Fig_X}. The refined redshift coverages, sizes, and centers are similar to their preliminary values. The C2 center agrees well with its X-ray center. For C1, there is an offset of $7\arcmin$ between its optical center and X-ray center. Such offsets have been reported in previous studies. {We will discuss this in Subsection \ref{offset}. We also calculate the galaxy number density profiles from the center to 3 Mpc, in a step size of 0.2 Mpc, and compare them with the X-ray luminosity profiles (Figure \ref{fig:growth}). The two types of profiles are generally consistent.}

The radii of the two clusters C1 and C2 are {1.9 and 1.7 Mpc} (see also Table \ref{t1}). Their line-of-sight sizes {in redshift space}, derived from the redshift coverages, are {39 and 28} Mpc, respectively. The sizes along the lines-of-sight appear much larger than the projected sizes, presumably due to the redshift space distortion (or the Fingers-of-God effect) \citep{Zehavi2001}. We will discuss this in the next section. In the following analyses, we assume that the clusters are spherical. The numbers of member galaxies from SDSS and MMT for C1 and C2 are {87 and 52}, respectively.

Finally, we estimate the masses of the two clusters using two independent methods. The first method is based on the galaxy number overdensity $\delta_g$. We use 
\begin{equation}
M_{\delta}=\frac{4\pi}{3}\cdot r^3 \cdot \delta_g/b \cdot \bar{\rho} \ ,
\label{mdet}
\end{equation}
where $r$ is the radius of a cluster, $\delta_g$ is the number overdensity of galaxies in the cluster, $b$ is the galaxy bias, and $\bar{\rho}$ is the average density of the universe.  We use the galaxy bias values from \cite{Crocce2016}, and take the first order for simplicity. The bias values are 1.07 and 1.06 for C1 and C2, respectively. In order to derive $\delta_g$, we calculate the average galaxy number density based on  SDSS galaxies in the archive.  We choose a three dimensional (3-D) spherical volume centered at the position of [R.A.=13h20m00s, Decl.=30d00m00s] at $z=0.07$, with a radius of 60 Mpc. There are more than 7000 SDSS galaxies in this volume that are sufficient to provide a reliable background density. In the calculation of $\delta_g$, the galaxy number densities in the clusters are also based on the SDSS galaxies, i.e., the MMT galaxies are not used. {The overdensity-based masses $\rm M_\delta$ of the two clusters within $\rm R_{200}$ are $\rm 7.9\times10^{14}\ M_\odot$ and $\rm 5.7\times10^{14}\ M_\odot$, respectively. We also calculated $\rm M_{500}$, and the results are listed in Table \ref{t1}. These results suggest that the two clusters are large and massive clusters.}

{
In the second method, we use galaxy velocity dispersions in the clusters to estimate their masses. We follow the formula given by \cite{Munari2013}, 
\begin{equation}
\frac{\sigma_{1D}}{\rm km\cdot s^{-1}}=A_{1D}\cdot[\frac{h(z)\cdot M_{\sigma}}{10^{15}~\rm{M_\odot}}]^\alpha \ ,
\label{msig}
\end{equation}
where $\sigma_{1D}$ is the galaxy velocity dispersion along the line-of-sight towards a cluster, $\rm M_{\sigma}$ is the cluster mass, $h(z)$ is the Hubble constant at redshift $z$, and $A_{1D}=1170$ and $\alpha=0.36$ taken from \cite{Munari2013}. The $\sigma_{1D}$ values of the member galaxies are 587 and 399 km $\rm s^{-1}$ for two clusters, respectively. The derived masses are $\rm 2.1\times10^{14}\ M_\odot$ and $\rm 0.7\times10^{14}\ M_\odot$, respectively. These masses are much smaller than the previous $\rm M_\delta$ masses. We will discuss this in Subsection \ref{me}.
}

\begin{figure}[t]
\centering  %图片全局居中
%\subfigure[Cluster\_1]{
\includegraphics[width=0.4\textwidth]{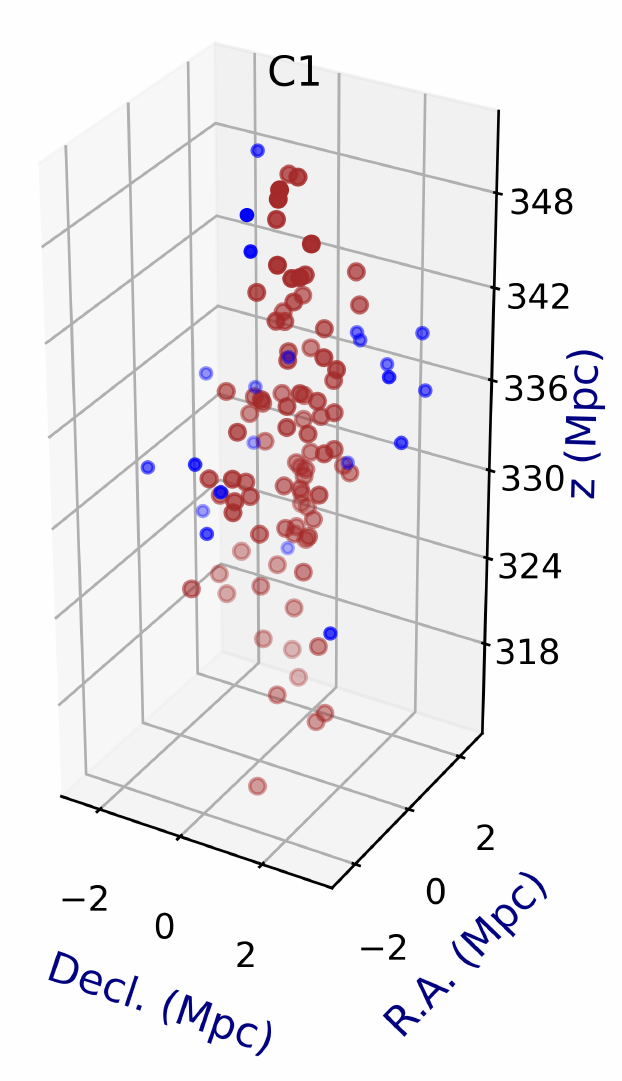}%}
%\subfigure[Cluster\_2]{
\includegraphics[width=0.4\textwidth]{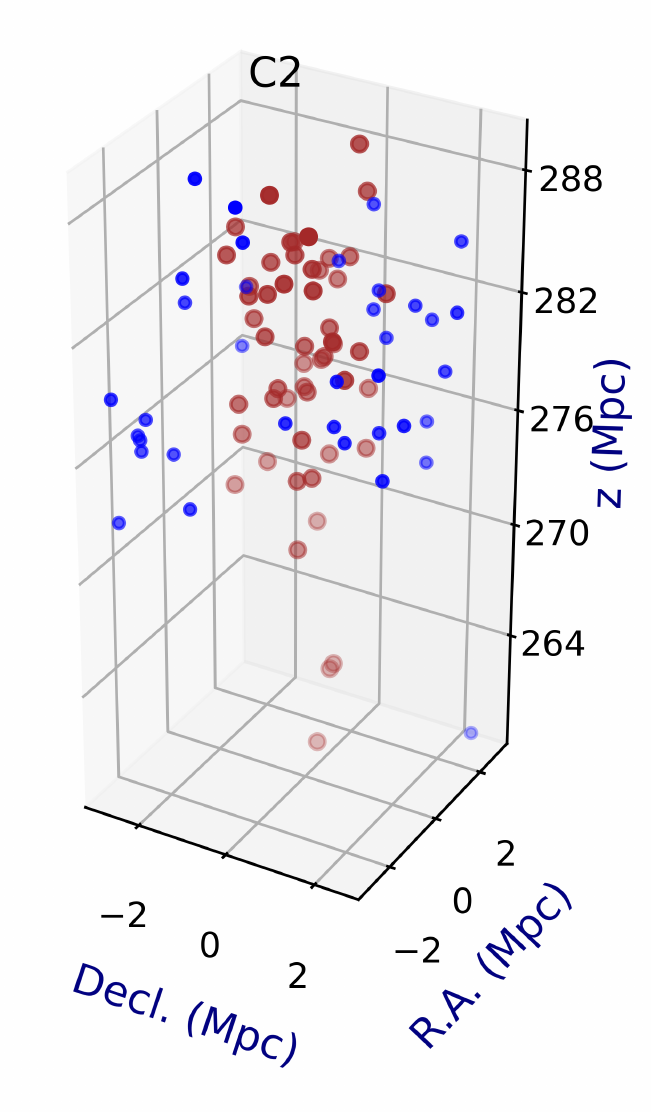}%}
\caption{
The 3-D distribution of the galaxies {in the clusters in the redshift space}. The Z-axis is the distance from us, derived from the galaxy redshifts. The brown circles represent the member galaxies in the two clusters. The blue circles represent surrounding galaxies. 
The coordinate origins of X-axis and Y-axis locate at the cluster center obtained from X-ray.
\label{Fig_3d}
}

\end{figure}

\section{Discussion}\label{discuss}

\subsection{{Offsets between the optical and X-ray centers}}\label{offset}
{
In section 3.2, we found that the cluster C1's center identified using its member galaxies is slightly different from its X-ray center. We search for the brightest member galaxies using the SDSS and look into this discrepancy here. For C2, there is no obvious offset between its optical and X-ray centers. The brightest galaxy with $g\approx 16$ mag in C2 is only 0.5$'$ off from the X-ray center. For C1, there are 3 brightest galaxies with $g\approx 16.5$ mag, and their distances to the X-ray center are 6.0$'$, 5.6$'$, and 2.4$'$, respectively. This suggests that C1 may have multiple bright central galaxies.
The X-ray position of C1 listed in Table \ref{t1} differs from its original RXGCC center, i.e., it has been corrected. The reason is that this RXGCC cluster suffers strong contamination from two nearby X-ray bright quasars. The two quasars are included in the ROSAT X-ray Source Catalog (RXS, \citealt{Voges1999,Voges2000,Boller2016}) and Half-Million Quasar catalog (HMQ, \citealt{Flesch2015}). The existence of the two quasars has affected the determination of the X-ray center in the work of \cite{xu_2021}. Thus, we masked out the two quasars in the RASS image and the newly determined X-ray center is shown in Table \ref{t1}. Due to the large ROSAT PSF, the residual X-ray emission from the two quasars may still affect the calculation of the cluster position.}

\begin{figure}[t]
\centering 
\includegraphics[width=0.7\textwidth]{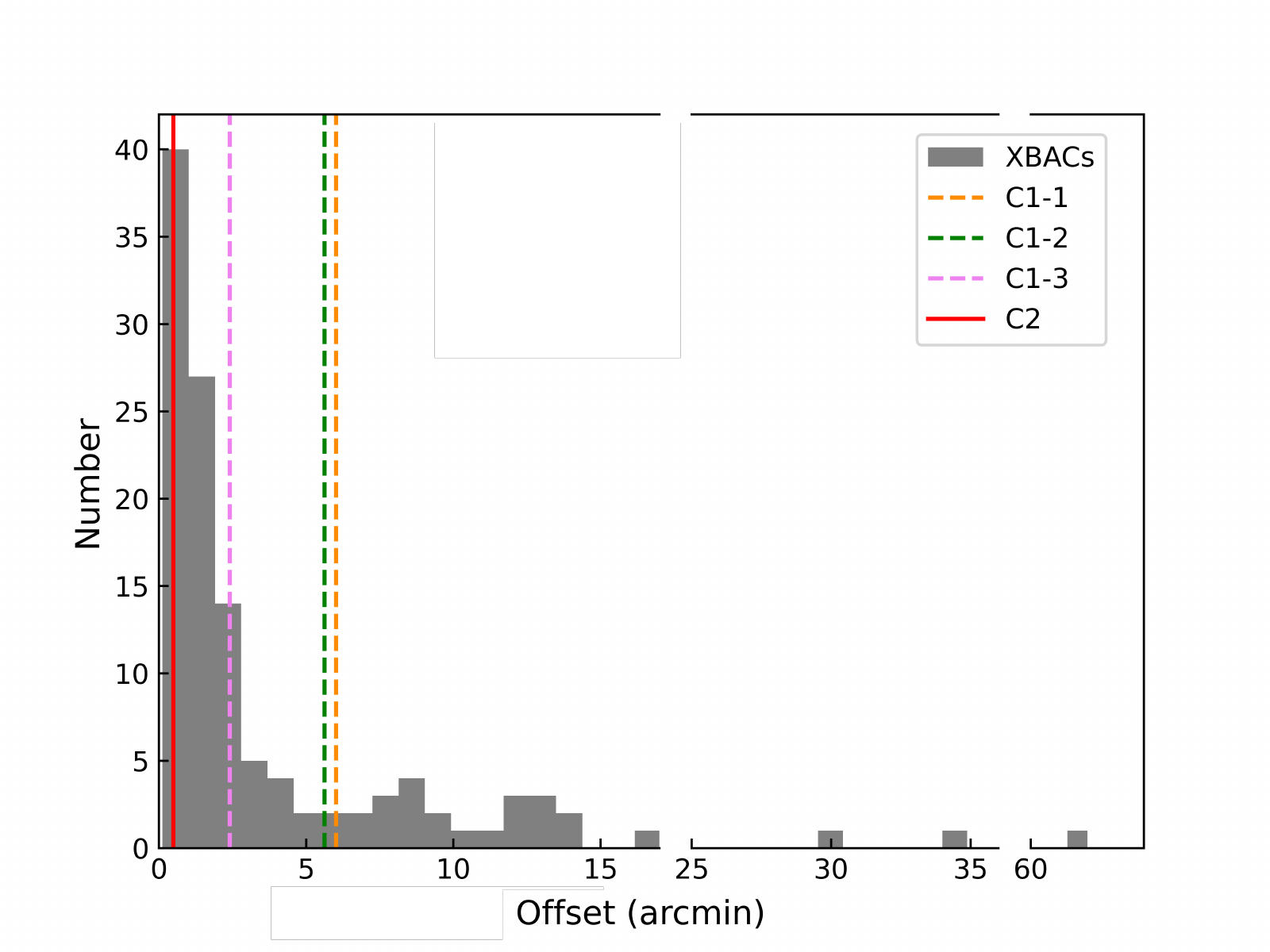}
\caption{{Distribution of the offsets between the optical and X-ray centers for 120 Abell clusters in the XBACs catalog. {The offsets between the brightest galaxies and X-ray centers for C1 and C2 are shown as the solid magenta line and the dotted red line, respectively.}}}
\label{Fig_offset}
\end{figure}

{
We further check the X-ray positions of optically bright clusters taken from the X-ray-brightest Abell-type clusters of galaxies (XBACs catalog, \citealt{Ebeling1996}). The XBACs catalog includes 120 X-ray bright Abell clusters, and their X-ray properties were measured with the ROSAT All-Sky Survey data. The corresponding optical positions of these XBACs clusters are obtained from \cite{Lauer2014}. Figure \ref{Fig_offset} shows the offset distribution of their optical and X-ray positions. The median and average offsets are 1.58$'$ and 4.34$'$, respectively, with quite a few offsets greater than 6$'$. Therefore, the large offset of C1 between its optical and X-ray positions is likely real.
}

\subsection{Sizes of the clusters}

In Section \ref{ic}, we mentioned that the line-of-sight sizes of the two clusters are much larger than their traverse sizes. In small scales, the galaxy distribution is elongated towards observers in the redshift space. This is due to the peculiar velocities of galaxies that produces random Doppler shifts. \cite{Zehavi2001} provided the following empirical formula to estimate such effect,
\begin{equation}
\xi_r =(s/s_0 )^{\gamma} \ \rm{,}
\label{fog}
\end{equation}
where $s$ is the traverse size of the cluster, $s_0=6.1\pm 0.2\ h^{-1}\rm{Mpc}$ is the critical size, $\gamma=-1.75\pm 0.03$ is the power-law index, and $\xi_r$ is the ratio of the line-of-sight size to the traverse size. They included more than 29,300 galaxies at redshift between 0.019 and 0.13, which are similar to the redshifts of our clusters. {Based on this formula, we find $\xi_r=4.3$ and 5.2 for the two clusters in this work. The observed $\xi_r$ values for the two clusters are 10.3 and 8.2, respectively. }They are generally consistent with the prediction, given that they could have been overestimated from the redshift distributions.

{
The two clusters show relatively diffuse profiles in the optical and X-ray. We use the concentration parameter from NFW Model \citep{Bullock2001} to describe the profile. In a hierarchically clustering dark matter halo, we have the following density profile
\begin{equation}
\rho(r) =\frac{\rho_s}{(r/r_s)\cdot(1+r/r_s)^2} \ \rm{,}
\label{nfw}
\end{equation}
where $r_s$ is the characteristic inner radius, and $\rho_s$ is the corresponding inner density. The concentration parameter is defined as $c_\delta=r_\delta/r_s$, where $\delta$ refers to the overdensity. We fit the theoretical profile to the observed number density profiles in a radius range from 1 to 2 Mpc (Figure \ref{fig:growth}). The observed number density here contains both MMT and SDSS galaxies. Poisson errors are used as uncertainties in the fitting. The best fits are {$r_s=0.58\pm 0.22$ Mpc and $\rho_s=47\pm 38$ Mpc$^{-3}$ for C1, $r_s=0.92\pm 0.41$ Mpc and $\rho_s=12\pm 11$ Mpc$^{-3}$} for C2. The associated $\rm R_{200}$ and $\rm R_{500}$ are consistent with the values derived in Subsection \ref{ic}, with differences less than 0.1 Mpc.
}

{
From the above calculations, we find that the concentrations $c_{500}$ (i.e., $c_\delta|_{\delta=500}$) for the two clusters are about 2.0 and 1.1, respectively. These values are much smaller than those in typical clusters ($4\sim5$) with similar masses \citep{Andreon2019}. However, they are consistent with a X-ray diffuse and weak cluster CL2015 \citep{Andreon2019} that has $c_{500} \approx 1.5$. This suggests that the clusters in this study with low concentrations are much more diffuse than typical clusters.
}

\subsection{Cluster masses}\label{me}
{
In Section 3.2 we measured cluster masses using three methods, including galaxy overdensities ($\rm M_\delta$), X-ray emission ($\rm M_X$), and galaxy velocities dispersions ($\rm M_\sigma$). The results were summarized in Table 1. There are significant discrepancies among the measured masses. We first compare $\rm M_X$ and $\rm M_\delta$. For C2, the two estimated masses within $\rm R_{500}$ agree with each other. For C1, however, the mass difference is roughly one order of magnitude. It is likely that the X-ray based mass was largely underestimated, due to the existence of the two nearby X-ray bright AGN, i.e., the AGN contribution could have been over-subtracted. In addition, such clusters are typically very weak in the RASS images, and thus it is difficult to obtain robust measurements of their X-ray emission. Furthermore, we cannot rule out that X-ray emission may underestimate the total mass in these diffuse clusters.
}

{
The discrepancies between $\rm M_\delta$ and $\rm M_\sigma$ in both clusters are also significant. This can be partly explained by our galaxy selection bias. In a cluster, brighter galaxies preferentially locate in the central region and thus tend to have relatively smaller velocity dispersions \citep[e.g.,][]{Biviano1992,Evrard2008}. In this work, the galaxies selected for the follow-up spectroscopy and for the analyses are relatively bright galaxies in the clusters. We test the selection bias in the two clusters using our data sample. We calculate $\rm M_\sigma$ separately using the SDSS galaxies and MMT galaxies. The SDSS galaxies are much brighter than the MMT galaxies on average. We find that the $\rm M_\sigma$ derived from the  SDSS galaxies is roughly $\sim 30\%$ smaller in either cluster. However, this is not enough to account for the large discrepancy between $\rm M_\delta$ and $\rm M_\sigma$.}

{
It is likely that the large discrepancy between $\rm M_\delta$ and $\rm M_\sigma$ is due to the diffuse nature of the two clusters. The formula (2) that we used is for normal clusters. As we have shown earlier, the clusters in this work are very diffuse and have much lower concentrations compared to normal clusters, so velocity dispersions may not reflect their total masses. More observations are needed to clarify it. 
}

\section{Summary}\label{sum}
{
We have spectroscopically confirmed two galaxy clusters selected from the RXGCC catalog \citep{xu_2018,xu_2021}. The catalog includes 303 ICM-detected clusters that were missed in other previous cluster searches using the same data. The two clusters show very extended profile and low-surface brightness in the RASS X-ray images. We carried out MMT Hectospec observations of their member galaxy candidates. Together with the SDSS archival data, we spectroscopically identified a large number of galaxies and derived the redshifts ($z=0.079$ and 0.067) of the two clusters. We used the member galaxies to measure cluster properties, including their central positions, sizes, overdensities, masses, etc. 
The central position of cluster C2 is consistent with its X-ray position. For C1, its optical and X-ray centers have an offset of about $7\arcmin$. Such offsets have been commonly seen in previous studies. The sizes of the clusters are about $R_{200} \approx 2$ Mpc. We used three methods to measure the masses of the clusters. We found that the masses derived from the galaxy number overdensities are about $\rm (6\sim8) \times10^{14}\ M_\odot$, suggesting that they are massive clusters. However, the masses based on the other two methods, galaxy velocity dispersion and X-ray emission, are apparently lower than the overdensity-based masses. The reason is still unclear, but this could be related to the diffuse features of the clusters.
}
If more such galaxy clusters are confirmed in the future, more accurate cluster mass function can be derived, which { will likely help to reduce the discrepancy} between the cosmological constraints obtained from galaxy clusters and from other methods such as CMB. 
  
\facilities{MMT (Hectospec), SDSS}

\begin{acknowledgments}

We also acknowledge support from the National Science Foundation of China (11721303, 11890693) and the China Manned Space Project (CMS-CSST-2021-A05, CMS-CSST-2021-A07).
We thank Y. Fu, Y. Guo, W. Liu, B. Weiner, J. Wu, and Z. Zheng for helpful discussions on observations and data reduction.
Observations reported here were obtained at the MMT Observatory, a joint facility of the University of Arizona and the Smithsonian Institution. 

\end{acknowledgments}

\bibliography{ms}
\bibliographystyle{aasjournal}



 \end{document}